# Switched systems and the logical foundation of circuit theory

*Emanuel Gluskin*

Academic Technological Institute, Holon 58102, Israel, *and* Electrical Engineering Department, Ben-Gurion University of the Negev, Beer Sheva 84105, Israel.
http://www.hit.ac.il/departments/electronics/staff/gluskin.htm

**Abstract**:  Switched (singular) systems become very common, which requires some revision of the conceptual basis of system theory.

## 1.  The ugly duckling approach to system theory

*Make things as simple as possible, but not simpler than that* (A. Einstein).  Yes, of course, but it is difficult in the very wide field of dynamic systems, fighting on so many fronts, to even define this "as possible".  Nevertheless, there is a clear motivation to consider the modern circuit situation because: (a) even good specialists, working in this or that subfield, make mistakes re classification of systems, and (b) our tools and arguments can help one to see analogies from other fields, which can contribute to making system theory a part of one's basic culture.  Isn't the absence of popular books in system theory/science intended for a wide public, something like "*Prelude to Mathematics*" by Sawyer, "*Geometry and Imagination*" by Gilbert and Con Fossen, "*Evolution of Physics*" by Einstein and Infeld, "*Physics for Everyone*" by Landau and Kitaygorodsky, etc., etc. regrettable?

   It is important to see that system theory is not just a branch of applied mathematics. For instance, the concepts of *input* and *output* are specific system concepts.  Take, e.g., a closed circuit, and connect to it two terminals anywhere, obtaining a 1-port. Since each branch of the (any) circuit also is a 1-port, you can create now a specific *fractal* circuit, by the recursive, step-by-step replacement of each branch of the existing 1-port by the whole structure of the 1-port.  Creation of such a fractal-circuit is not the usual mathematical approach to fractals since "port" is not a mathematical concept, and, if we mean not only the circuit graph, but also some not necessarily ideal elements, then even the basic system concept of "structure" is also not a mathematical concept.

   Thus the ugly duckling (i.e. physically and philosophically) opinion here is that system theory is an *independent science*, and *as such* deserves to have its own axiomatization and to become an organic part of one's general culture.  The time has came, it seems, even for a proper course, created both by specialists and teachers, to arise in the secondary school education, and the material given below may cause the reader to think even in this non-standard direction.

   Seeking a paradigm for what follows among the books of my childhood, I would mention "*Elementarmathematik vom hoheren Standpunkte aus*" by Felix Klein, where the "high standpoint" is *not* so much the mathematical technique, but rather the addition of some simple logic and common sense to the usual pedagogical positions.

## 2. It is not trivial at all to define the switched circuit, and Gustav Robert Kirchhoff would not be happy with the definition

Wishing to speak about switching (or sampling [1-4]; generally, *singular*) systems, let us follow [1-4], denoting the set of the instants $t_k$ of singularity of a *time-function* of a system as $\mathbf{t^*} = \{t_k\}$. If we switch an element of a system at $t = t_1$, then such a function either "jumps", or receives a sharp inflection, i.e. the derivative jumps at $t_1$, and the singularity can be expressed in jumps of higher order/derivatives too. Let a function denoted at $x_3(t)$ to jump; then such a function as $x_{13}(t) \sim \int^t x_3(\lambda)d\lambda$ receives respective jumps of its derivative, at the same instants. Different *mutually connected* functions can thus have different degrees of singularities, but in any case, some singularity has to appear in the system. This presents a problem because a function with a singularity has an infinite frequency spectrum i.e. includes some frequencies for which the lumped circuit condition (the wavelengths must be much larger than the circuit's physical dimensions) is violated. This means that the switched circuit radiates electromagnetic energy, which it is not at all simple to take into account mathematically in circuit analysis, and the not quite simple Maxwell equations that are the base of the Kirchhoff's equations have to be applied … .

   Yes, the dynamics of a switched system can be compared with nice music played using a stretched disk!

   That the fact that switched circuits are known as "noisy" circuits is thus associated with the fact that somewhere close (in time) to the switching instants they are not lumped circuits, and, strictly speaking, we cannot include, as we do below, $\mathbf{t^*}$ in the formulae of *such* a circuit, which are claimed to describe the processes at any $t$.

   Let us thus agree that our formulae will describe the process not very closely to the $t_k$, even though we shall not try to develop here any estimate of how "not close" it is. We thus follow the M.S. student, Claude Elwood Shannon, who was not anxious to ignore such physics problematicity in his work [13] showing that the analysis of *switched* (relay) circuits is reduced to the application of Boolean algebra, and thus starting the real development of computers. Well, … who would not agree that the important piece of advice to be given to a young scientist is: *"Do not be afraid to make mistakes!"* ?

   The natural frame of the subject forces us to speak not only about switched systems but also about some classical *analytical* systems. However the main line of thought passes through switched systems.

## 3. Switching systems in the context of linearity and nonlinearity and the role of the instants of singularity of time-functions

When having some it's elements switched, any physical/hardware structure, relevant to the mathematical description in focus, is changed in time. If this change is *completely prescribed*, which requires, in particular, $t_k$ to be given *a priori*, we deal with an LTV (linear-time-variant) system, and if this change (and $t_k$) cannot be prescribed, i.e. is dependent on the processes being developed in the system, then we deal with an NL (nonlinear) system. Because of the switchings, LTI (linear time-*invariant*) systems are *irrelevant* here. However, the LTI *elements*, usually constant capacitors or resistors, are the main objects of the switching.

When the instants of singularity **t\*** are prescribed, we write **t\***(*t*), and when they are dependent on one or several unknown state-variables $\{x_p\} \in \mathbf{x}$ to be found in the same system, then we write **t\*(x)**. Of course, $t_k$ are finally just some *numerical* parameters, and the notations **t\***(*t*) and **t\*(x)** not so much mean analytical dependences, but rather the very facts that $t_k$ are *defined* by some properly classifiable functions. That is, the point is more *informational* than analytical; we just ask the designer of a device, or one who analyzes it, to *observe* (or *detect*, as in [2]) any such influence.

Symbolically written, the notation **t\***(*t*) means a *map*:

$$f(t) \rightarrow t_k \qquad (1)$$

for a *given* (prescribed, fixed) function *f(t)*, and the notation **t\*(x)** means

$$x_p(t) \rightarrow t_k \qquad (2)$$

for a state-variable $x_p \in \mathbf{x}$.

Let us consider how switching of the usual elements can create either LTV of NL system. Most simply, the influencing function, which can be named the "informative" function, or the "trigger", is an input of a comparator. See Fig. 1.

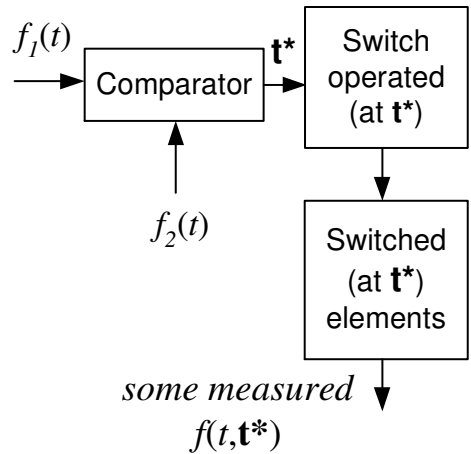

Fig. 1: Generation of switchings instants **t\***(.) = $\{t_k\}$. In the NL case, i.e. that of **t\*(x)**, $f_1$ or/and $f_2$ are some $x_p$, though the influences of a state-variable can be indirect, of course, i.e. $f_1$ can be *dependent* in some way on a state-variable. The circuit situation in [6] can be included in the latter case. If neither $f_1$ nor $f_2$ is connected with any $x_p$, then we have an LTV system, **t\*** = **t\***(*t*). It is most natural to take $f_2$ as a known reference function, leaving for $f_1$ to define the nature of the switching operation, it to be linear or nonlinear.

Thus, our $\{t_k\}$ = **t\*** are some level-crossings (either known a priori, or not) of some time-functions existing in the system. Figure 2 illustrates this action of the comparator.

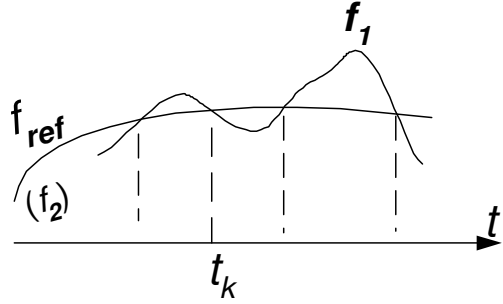

<u>Fig. 2</u>: The map $f(t) \rightarrow t_k$, or $x_p(t) \rightarrow t_k$, depending on *what $f_1(t)$ is*; $f(t)$, or $x_p(t)$.

Thus, in terms of **t\***($t$) and **t\***(**x**), works [1,2] began the classification of switched systems, composed of LTI elements. "Began", because if we deal with non-autonomous systems having some *inputs* $\{u_r(t)\}$ = **u**($t$), then this classification also includes the case of **t\***(**u**). We mean that $f_1$ in Figs 1 and 2 can also be one of the $u_r(t)$.

All three cases of **t\***($t$), **t\***(**x**), and **t\***(**u**), are relevant to *one* equation that expresses, via **t\***(.), the fact that in each case we deal with a *switched* (singular) system:

$$d\mathbf{x}/dt = [A(t, \mathbf{t}^*(.))]\mathbf{x}(t) + [B(t, \mathbf{t}^*(.))]\mathbf{u}(t) . \qquad (3)$$

It is easy to include **t\*** into the matrices. Just start with an LTI system, having, say, a constant resistor, thus having [A($R$)], and then begin to switch (change) $R$ at some instants, using a proper "informative function". It thus becomes obvious that $t_k$ should be included in the analytical expressions of the time-functions that describe the switched elements (say, $R(t,t_k) = R_1 u(t_k-t) + R_2 u(t-t_k)$, where u($t$) is the unit step), and then in the matrices, by means of the differences $\{t-t_k\}$, i.e. as some time-shifts. Any such process is developed as some "sewed" analytical pieces. *However*, the notation [A($t$,**t\***)] remains most suitable and relevant since we *observe* **t\*** separately, choosing between (1) and (2), and only this is our focus here. Thus, we shall continue to speak about $\{t_k\}$ and not $\{t-t_k\}$. Indeed, it would look just odd to replace $f(t) \rightarrow t_k$, or $x_p(t) \rightarrow t_k$, by $f(t) \rightarrow t-t_k$, or $x_p(t) \rightarrow t-t_k$ !

### 4. The main equational cases

According to the cases of **t\***($t$), **t\***(**x**), and **t\***(**u**), the respective forms of the basic equation (3) are (we just substitute the respective **t\***(.) into (3)):

$$d\mathbf{x}/dt = [A(t)]\mathbf{x}(t) + [B(t)]\mathbf{u}(t) \qquad (4)$$

$$d\mathbf{x}/dt = [A(t,\mathbf{x})]\mathbf{x}(t) + [B(t,\mathbf{x})]\mathbf{u}(t) \qquad (5)$$

and

$$d\mathbf{x}/dt = [A(t,\mathbf{u})]\mathbf{x}(t) + [B(t,\mathbf{u})]\mathbf{u}(t) , \qquad (6)$$

which can respectively be:

$$d\mathbf{x}/dt = [A(t)]\mathbf{x}(t), \qquad (4a)$$

$$d\mathbf{x}/dt = [A(t,\mathbf{x})]\mathbf{x}(t), \qquad (5a)$$

and

$$d\mathbf{x}/dt = [A(t,\mathbf{u})]\mathbf{x}(t). \qquad (6a)$$

How can we check which equation is linear, and which nonlinear? Of course by using the two (scale and/or add) simple tests of linearity! The *scaling test* is sufficient here:

$$k\mathbf{u}(t) \rightarrow k\mathbf{x}(t)$$

(and $\mathbf{x}(t) \rightarrow k\mathbf{x}(t)$ for the autonomous systems (4a) and (5a)), and it is seen (check!) that among (4)-(6a) only (4) and (4a) are linear.

## 5. A question regarding nonlinearity of (6)

One might find the fact that equation (6) (or (6a)) does not pass the test of linearity a bit strange because the input $\mathbf{u}(t)$ should be *given (prescribed)*, and thus in (6), *just as in* (4,) the matrices are *known*, i.e. it seems that the system operation (switching) is prescribed in both cases. Thus, the fact that (4) is a linear equation (system), while (6) nonlinear, may seem to be paradoxical; both should be linear, one says.

In order to better understand this pedagogically important point and the very concept of the "**u**-nonlinearity", one should note that in system theory (in fact, also in mathematics, but in "systems" this is really important) the concept of "given function" has a double meaning.

In (4), some fixed functions, defined by the producer of the device, and included in [A(t)], are given *and they can not be changed*. Contrary to that, in (6), $\mathbf{u}(t)$ (or any its component) denotes a given *set of functions*, from which we each time pick what we need, and which includes, in particular, a linear (sub)space allowing us to perform the test of linearity for (6) by taking linear combinations of the functions.

Of course, a certain given function that can interest one just by its waveform, i.e. the case of (4), cannot be equivalent to the use in (6) of a whole space of functions in which some *operations* are defined. In addition, any "input" is an interface of a system with the external world, and by allowing one to change (in a non-prescribed way) the input, we in fact turn the system into a subsystem of some wider unknown system.

This distinction between the two meanings of "given/known" functions, relevant to the cases of (4) and (6), eliminates the seeming "paradox". Since the circuits' situations are different, the conclusions regarding linearity or nonlinearity need not be similar, and one can rely on our use of the standard test of linearity without any worry!

Observe that in this whole argument the *informational (logical) aspect* of the problem (and not any analytical details) is dominant!

Despite the phenomenological clearness of the point, the things should be a bit new for a reader who might be amazed, e.g., by the question of whether or not the equation

$$\frac{dx}{dt} + a(t)x(t) = u(t), \qquad (7)$$

or, rather

$$\frac{dx}{dt} + a(t)x(t) = 0 \qquad (7a)$$

(having the solution

$$x(t) = x(0)e^{-\int_0^t a(\lambda)d\lambda}$$

that shows that the map $a(t) \to x(t)$ is *multiplicative*), -- is linear. The answer depends on whether $a(t)$ is just a fixed function in the sense of (4), or a system-input that must be changeable, as in (6).

Of course, many always assume that the right-hand side of an equation is the reserved place for the "input". However one can also see that "right-hand side" is a *physiological* concept having no relation to mathematical rigor. What is really relevant, is that *the very definition* of the system for which the equations have to be written requires definition of the inputs that in the modern systems (e.g. vision chips) can be very numerous, and we think that a mathematician wishing to contribute to system theory should not ignore this.

## 6. Another funny example

Another very simple example, perfectly illustrating the change in the equational classification introduced by the system outlook, is as follows. Compare the two equations including a linear operator $\hat{L}$ and some *known* functions, $f_1$ and $f$:

$$\hat{L}x(t) + Af_1(t) = Bf(t) \qquad (8a)$$
$$\hat{L}x(t) + f_1(t) = Bf(t). \qquad (8b)$$

Here, $A$ and $B$ are scaling parameters symbolizing (such *agreement* is a purely system matter, or, rather, system *reality*, and for a pure mathematician this item can be unexpected …) that a function having such a scaling parameter is an *input* of a system. (And if so, then not only the scaling, but also addition, or changes performed by any other means, can be applied to such a function.) That is, in (8a), $\mathbf{u} = (f_1, f)^{\mathrm{T}}$ or, simpler, $\{f_1, f\}$, and it is a linear equation

$$\hat{L}x(t) = Bf(t) - Af_1(t)$$

for which $\mathbf{u} \to k\mathbf{u}$ (i.e. $\{f_1, f\} \to \{kf_1, kf\}$) *is* followed by $x \to kx$. Contrary to that, in (8b), i.e. in

$$Bf(t) = \hat{L}x(t) + f_1(t),$$

$f_1(t)$ is *fixed, given by the producer of the system*, and this equation (system) is *not* a linear, but an *affine* one, i.e. of the type $y = ax + e$ with '$e$' fixed, for which $y \to ky$ and $x \to kx$ do *not* coexist, since the free term cannot be changed. In some sense, $f_1$ acts as a hardlimiter, which is very far from anything linear. Alas, despite the presence of the magically impressing linear operator, the form $\hat{L}x(t) + f_1(t)$ is *not* a linear one.

Concluding this, unexpected for one, but *undoubtedly theoretically important* nuance of the equational classification, let us ask the following question that is rhetorical not because there is an obvious answer, but because it is too early for the answer to appear. *Are the concepts of input and output just some system, or also some "mathematical" concepts?* To what extend can the physical and the *logical* definition of a system force a mathematician to complete his conceptual vision originating from the old mechanical problems related to systems of simple structures, with one or two simple inputs, which were/are never treated using the concept of **u**-system ([A(**u**)]-system) because this is not actual for such problems?   We meet here some natural inertia of the well-developed standard thinking, associated with the automatic use of the powerful apparatus of differential equations.  This power is highly respected, but contrary to the common opinion, the classification is not closed.

Meanwhile, let us conclude that in system theory the classification of systems as linear or nonlinear should be a part of its own axiomatization that is not yet sufficiently developed.

## 7. Not necessarily switched systems

Since **t\*** is not seen in the "final" equations (4-6), the obtained classification: {(4) – linear, (5-6) – nonlinear} does not necessarily relate to singular systems.   For analytical systems for which [A($t$)] or [A(**x**)] are obtained not via **t\***(.), this classification can be accepted by many, with the reservation employing the equational "normal-form" $d\mathbf{x}/dt = \mathbf{F}(t,\mathbf{x}(t),\mathbf{u}(t))$. Namely, if $\mathbf{F}(t,\mathbf{0},\mathbf{u}(t)) \neq 0$, then passing to the constructive form (5) or (5a) is unnatural, and the normal form can be preferred. However this is the only limitation on the dynamic systems, for our "structural" outlook to be applied, and one sees that the very point of linearity or nonlinearity remains in force even when the state-state ([A]-[B]) description is not natural.

However also in this general scope it is good to come to (4,5) via switched (singular) systems because of:

 (a) the drastic change in the whole scope of nonlinear systems, caused by the increased use of switched systems;

 (b) the fact that the *amplitude-phase relation is a well-known "classical" feature of a nonlinear system*, which is perfectly seen via switched systems.  (Indeed, since any $t_k \in \mathbf{t^*}$ actually appears as a time-shift of a singular time-function, it can be seen as related to the "phase" of a function, while **x** in **t\***(**x**) and **u** in **t\***(**u**) include amplitude parameters;  thus, in view of the old well-known argument, our equational classification is just natural!)

In order to consolidate the things, explain why among the following expressions (a)-(h), including the function *signum* (sgn), only (a)-(c) are linear.

(a) $x_p \sin\omega t$,  (b) $x_p \text{sgn}[\sin\omega t]$,  (c) $x_p \text{sgn}[\sin\omega(t-t_k)]$  ($t_k$ fixed); (d) $x_p^2 \text{sgn}[\sin\omega t]$;
(e) $\text{sgn}[\sin\omega(t-t_k(x_p))]$;  (f) $x_p^2 \text{sgn}[\sin\omega(t-t_k(x_p))]$,  (g) $x_p \text{sgn}[\sin\omega(t-t_k(x_p))]$,
(h) $\text{sgn}[(x_p(t)]$.

## 8. Why designers of electronic systems need this outlook?

This is because the connection between **t\***(**x**) and [A(**x**)] (i.e. A[**t\***(**x**)]) simply explains that the "**x**-control" of switching (sampling) means nonlinearity. It is known for many power specialists that a feedback in power electronics can improve stability of operation of a device, making, at the same time, some of its operational characteristics nonlinear. However, the output variable used in such a feedback is only *one* of the possible $x_k$, while we speak about *any* state-variable. If in a system of the 100$^{th}$ order, $x_{72}(t)$, which is the (initially unknown) voltage on a capacitor placed in the electrical scheme far from the switched element(s), operates the switch(s), then the system is nonlinear.

In [1-4] some examples are given, but one can easily invent many such examples, and however complicated.

## 9. Why theoreticians need this outlook?

There are several reasons for that. Let us start from two important works of outstanding specialists in chaotic systems, [5] and [6].

A point in [5] should be corrected. The capacitor involved in a circuit (that of Fig. 2 there) is said to be "nonlinear", but is defined as (characterized by) $C(t)$, i.e. as a function that everywhere in the sequel of [5] is approached as known (prescribed). In such a situation, the whole system is of the type [A($t$)], i.e. (4)

Thus, contrary to [5], this is, *by definition*, a linear, and not a nonlinear system, and contrary to the impression expressed in [5], there is nothing surprising in the passing a linearity test (see equations (2) and (1) in [5]) by the circuit.

The tendency in the sequel of [5] to almost completely "jump over" LTV systems, directly opposing nonlinear systems to LTI systems, also causes worry, in particular because LTV systems are a wide class of dynamic systems, which is important for engineering (e.g. linear switched capacitor circuits) and for physics (e.g. Schrodinger's equation, parametric resonance equation, etc.), and should not be ignored in a tutorial on the features of nonlinear systems, such as [5].

A close comment is presented in [7] devoted to [6]. Work [7] states, regarding the understanding of the mathematical situation in [6], that *the switching between some linear systems* in [6] is a *nonlinear operation* simply because in [6] the switching instants $\{t_k\} \equiv \mathbf{t}^*$ are dependent on the state-variables $\mathbf{x}(t)$ obtained in the whole system, $\{t_k(x_p)\} = \mathbf{t}^*(\mathbf{x})$. This fact causes, as we already know, [A($t$,$\mathbf{t}^*$)] = [A($t$,$\mathbf{t}^*$($\mathbf{x}$))], or, simpler written, [A($\mathbf{x}$)].

Observe that in [6] $\mathbf{t}^*(\mathbf{x})$ is created differently from the way employed here. Namely, the map $x_p(t)$ → $t_k$ is associated in [6] not with detection and the use of some level-crossings of the functions $x_p(t)$ by themselves, which are $t_k$, but with some more complicated mathematical requirements on $\mathbf{x}$, influencing $\mathbf{t}^*$. Since this distinction does not violate or hide the point of the $\mathbf{t}^*(\mathbf{x})$-nonlinearity, it just makes work [6] very interesting from the positions of our research.

We think that the axiomatically important point of Section 5, i.e. the [A($\mathbf{u}$)]-systems, also should interest theorists, and the next paragraph explains an even more important point, to be seen by theorists and designers.

## 10. Do not make this mistake!

Sometimes one says that all switched systems are nonlinear. This is obviously wrong, since linear switched capacitor circuits (e.g. the famous modeling of a resistor by means of capacitor and two switches) are well known, and parametric resonance in an *LC* circuit is often demonstrated using prescribed two-levels sharp switching of the capacitor ($C_1 \leftrightarrow C_2$), which is described by an equation of the type $x'' + a(t)x = 0$ with a prescribed (*fixed, as in* (4a)) singular $a(t)$. Denoting $x_1 = x$ and $x_2 = x'$, one easily rewrites this second-order scalar equation as a matrix-equation (5a).

If one takes a *smooth a(t)* and well approximates it by small steps symbolizing some switchings, then the statement that all switched system systems are nonlinear can even lead one to the absurd idea that no LTV systems exist at all!

Where are the roots of this mistake? The point is that one thinks that any switched unit has a "characteristic", and combines this with the fact that any linear piece-wise characteristic is a nonlinear characteristic, i.e. that any given inflection means nonlinearity. The latter is true, of course, but the assumption that each switched unit has a "characteristic" with some *fixed* inflection(s), is wrong.

In the context of switched systems there are two possibilities. Observing switchings via inflection of a line (expected to be a "characteristic") in the plane of some state variables, $x_1$ and $x_2$, we either *have a stable result of the observation*, and thus really a certain characteristic of the switched unit which *is* nonlinear, or the inflection point appears, with the sequential switchings, at different points of the plane, and the switched unit under study does not have any "characteristic". The latter case (Fig. 3) can give an LTV system.

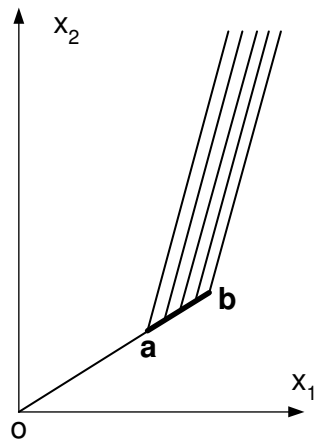

Fig. 3: When $t^* = t^*(t)$, there is no certain "characteristic" of a switched unit in terms of the state-variables because the inflection point is not the same point $(x_1,x_2)$ at each switching instant $t_k$, but for an NL unit, we do have a fixed inflection point in this plane, and a certain piecewise-linear, *nonlinear*, characteristic exists. The condition **ab << oa** is relevant for analysis of closeness of the NL and LTV cases, while the ratio of the slopes is a factor, associated with clearness (effectiveness) of the switching, in both cases.

It is important to consider such a figure because it shows that for switched systems, closeness of the LTV and NL cases can be defined, in principle, by parameters of two types: first of all by the value of **ab**, and then by the ratio of the slopes.

However, in both LTV and NL cases, we can observe singularity of a *time*-function (some $x(t)$), and our advice is to *directly work*, as much as possible, with time-functions (and not with characteristics), combining the observation of the singularities with the discussed "**t\***(.)-principle". Then there is no place for such a mistake.

## 11. Some examples

Figure 4 gives examples of an LTV and an NL systems for the phenomenological outlook developed

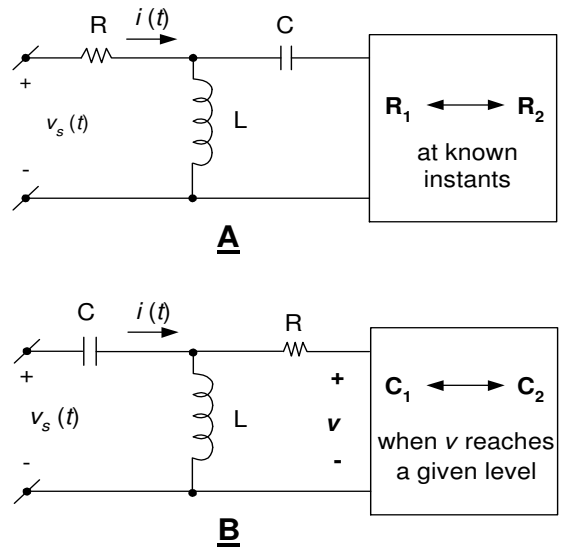

Figure 4: Circuit **A** is linear, and **B** nonlinear. One can check this, e.g., by study the map $v_s(t) \rightarrow i(t)$; is it linear convolution or not? In case **B**, the switched unit has a (nonlinear) "characteristic", and in case **A** it does not have. Case **B** is of the type [A(**x**)] ($v$ is our $x$), and case **A** is of the type [A($t$)], i.e. in case **A** we have **R**($t$), and in case **B**, **C**($t,v$), obtained as **C**($t$,**t**\*($v$)).

A more detailed example is shown in Fig. 5

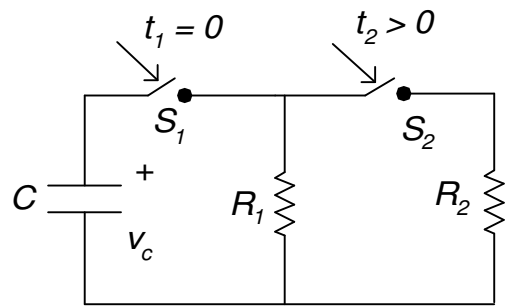

Fig. 5: $S_1$ is closed at $t = 0$, and $S_2$ is then closed *either* at any prescribed/independent moment, or when $v_c$ reaches a certain level $v_{cr}$. In the first case, the circuit is linear (LTV), and in the second NL.

Let the initial voltage on the capacitor be $v_o$. Since this voltage is not given to us by the producer of the capacitor, $v_o$ is as legitimized an "input" as any function generator connected to any port, just relates to ZIR and not to ZSR. One even recalls that if a circuit is presented in the Laplace-variable domain, then such similarity between the two kinds of inputs becomes an "official" one; the initial values become some sources. Thus, initial conditions should often be seen as a part of a generalized input that includes also the generators' functions, and, in fact, for a linear circuit, linearity of the *complete* circuit response (ZIR + ZSR) is with respect to *such* an input. After begging the reader's pardon for recalling what he well knows from the school bench, let us, as the point, *check the possible linearity of the circuit by observing whether or not the voltage on the capacitor is directly proportional to $v_o$, at any moment $t>0$.*

Under this criterion we shall analyze two states/processes that are distinct regarding the operation of switch $S_2$. In the first, the moment $t_2$ of closing $S_2$ is prescribed, and in the second, $S_2$ is closed when $v_c(t)$ reaches a certain given "critical" voltage level $v_{cr} < v_o$. Our phenomenological outlook clearly states that the first process is linear (LTV), and the second nonlinear, but the analytical details may be of a methodological interest, especially because we shall finally suggest related problem as home work.

Since for $t < t_2$ we have an LTI circuit in both cases, let us focus only on the process at $t > t_2$ i.e., after the crucial switching occurred.

For both cases, we can write for $t > t_2$ ($v_c(t_2)$ is initially parenthesized in (9)):

$$v_c(t) = \left( v_o e^{-\frac{t_2}{\tau_1}} \right) e^{-\frac{t-t_2}{\tau_2}} = \left( v_o e^{(\frac{1}{\tau_2}-\frac{1}{\tau_1})t_2} \right) e^{-\frac{t}{\tau_2}}, \quad t > t_2, \quad (9)$$

where $\tau_1 = R_1 C$, and $\tau_2 = R_1 R_2 (R_1+R_2)^{-1} C < \tau_1$.

In the case of $t_2$ prescribed, independent of $v_o$, we have from (9) $v_c(t) \sim v_o$, i.e. linearity.

In the second case, $t_2$ is defined by the equation,

$$v_o e^{-\frac{t_2}{\tau_1}} = v_{cr}$$

i.e.

$$t_2 = t_2(v_o) = \tau_1 \ln \frac{v_o}{v_{cr}}, \quad v_o > v_{cr},$$

and from (9)

$$v_c(t) = \left( v_o e^{(\frac{1}{\tau_2}-\frac{1}{\tau_1})\tau_1 \ln \frac{v_o}{v_{cr}}} \right) e^{-\frac{t}{\tau_2}} = v_o \left( \frac{v_o}{v_{cr}} \right)^{\frac{\tau_1}{\tau_2}-1} e^{-\frac{t}{\tau_2}} \sim v_o^{\frac{\tau_1}{\tau_2}}, \quad t > t_2.$$

Since $\tau_1 \neq \tau_2$, the dependence of $v_c(t)$ on $v_o$ is nonlinear, and the circuit is nonlinear.

Of course, for $R_2 >> R_1$, when $\tau_1$ becomes close to $\tau_2$, the nonlinearity becomes weak. Observe also that since $\tau_1/\tau_2 > 1$, this NL process is *more sensitive* to changes in $v_o$ than the LTV process.

This problem can be made more interesting by, e.g., including more switches and resistors becoming connected in parallel, one after another, at the smaller and smaller critical voltage levels. After obtaining the precise solution in each time interval, it can be curious, for instance, to use the conditions of fixed total charge $Cv_o$, and the total energy $Cv_o^2/2$, in order to set some conditions on the $\{t_k(v_o)\}$. However, since our main focus is not the analytical, but the *logical* one, we shall leave any such circuit-variation to the interested reader, and return to the most important heuristic line.

## 12. Why just electrical circuits? Introductory comments on two mechanical analogies

For our primary educational and heuristic purposes, let us consider two relevant mechanical systems, both found around us and both very basic. The first one, -- ensemble of colliding particles, say molecules of the air -- appears to be, in our notations, a **t\*(x)**-system, and the second one – a liquid flow, -- is shown to be an *analytical* [A(**x**)]-system. For the latter example, one should recall some basic notations of the vector analysis, which are taught, e.g., in basic courses of electromagnetic fields, which all EE students have to pass through. For one aquainted with the continuity equation $\frac{\partial \rho}{\partial t} + div\, \vec{j} = 0$, the equation $\nabla \vec{v} = 0$ (or $div\, \vec{v} = 0$) for velocity of an incompressible liquid is clear, and such scalar differential operators as

$$\Delta \equiv \nabla\nabla \equiv \frac{\partial^2}{\partial x^2} + \frac{\partial^2}{\partial y^2} + \frac{\partial^2}{\partial z^2} \quad \text{and} \quad \vec{v}\nabla \equiv v_x\frac{\partial}{\partial x} + v_y\frac{\partial}{\partial y} + v_z\frac{\partial}{\partial z} \quad \text{are also well}$$

known.

Operator $\vec{v}\nabla$ will be especially important in the description of the flow. Is it linear or nonlinear? *This depends on the function on which it acts*. If it acts on **v** or a function connected with **v**, then the result is as the nonlinear term [A(**x**)]**x** in (5) or (5a), i.e. the operator is nonlinear. But if **v**(*t*) were to be prescribed, and the operator were to act on another function (a kind of $x_p$), then it is a linear operator.

## 13. An ensemble of colliding particles, and the "thermalization" resulting from a t\*(x)-nonlinearity

Let us observe that for an ensemble of many colliding particles, the chaotic movement of the particles is obtained because of the nonlinearity of the "switching" type. The instants of the collisions of the particles (Fig. 6) are the points of the singularity. They *depend* on the trajectories $r_i(t)$ of the particles (our "state-variables"), because they are defined by equalities of the type $r_i(t) - r_j(t) = 0$, $i \neq j$, (which are our "comparator"-actions), i.e. this is a **t\*(x)**-nonlinearity.

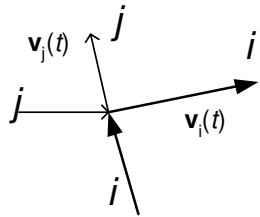

Fig. 6: Two colliding particles, number *i* and number *j*. On the one hand, the instants of collisions, **t\*(r)**, are defined by the trajectories $\{\mathbf{r}(t)\}$ and more directly by the velocities $\{\mathbf{v}(t)\} = \{d\mathbf{r}/dt\}$, and, on the other hand, these instants are the time-shifts of the spike-forces acting on the particles. Thus, the dynamic (Newton's) equations must be nonlinear: $d\mathbf{r}/dt = \mathbf{F}(t, \mathbf{t^*(x)})$. More details are found in [4].

This consideration can even lead one to the following *assumption*:

*Any ensemble of colliding particles where a chaotic distribution of the movement is obtained is a nonlinear system with a kind of nonlinearity which is close to that found in nonlinear switched systems, i.e. the **t\***(**x**)-nonlinearity.*

We say "*close to*" because real particles, e.g. molecules, are not rigid balls, and the collisions are not momentary. However, as was explained in the Section 2, for an electronic circuit too the precise $t_k$ appear as an equational fiction, because of the physical reality!

Of course, it is important for the argument of nonlinearity that we spoke in terms of detailed dynamic description. For applicational purposes, ensembles of many particles are (have to be) treated statistically. Then, the hard problem of solving the numerous equations disappears, but the important nonlinearity of the collision processes is already not seen.

It is interesting to note that some modern circuit theorists study *statistical features* of electronics chaotic systems. This tendency is natural and expected because, contrary to physics that is ready to study a phenomenon occurring once per thousand years, engineers think about simple applications, and each device, e.g. a chaos generator, is finally intended to be sold, and thus has to have some "structurally stable" specified characteristics. After system science covered the topic of stable oscillations, and limit cycles were "replaced" by smoothed chaotic attractors (obtained by means of however complicated bifurcations) nothing but the stability of the statistical properties of these attractors (and thus of the statistical spectral properties of the chaotic processes) could appear standing "on line".

These circuit-theory studies have to be compared/connected, however, with the classical physical statistical methods; at the least, the pedagogical side requires one to do this.

## 14. Liquid flow as an "[A(**x**)]-system" with obvious dependence of the "structure" on the "state"

Look at a liquid flow! (Fig. 7) Not knowing any hydrodynamic equation, just *thinking in system terms* (and we do want the "system thinking" to become part of one's general culture!), one sees that the *structure* of the flow considered as a "system" (that can have "inputs" and "outputs", etc.) is just the *same* velocity vector field **v**(*t*) -- an important *part* of the state-variables (our "**x**") that have to be found from the hydrodynamic equations. We say "*part* of the state variables" because there is also one more important state-variable, the pressure *p*, which is not directly seen.

That is, the dependence of the "structure" on "**x**" is inherent for the flow, and thus the situation is just as with an electronic [A(**x**)]-system. It becomes obvious that the hydrodynamic equations must be nonlinear, and, in particular, turbulence can be seen as a chaos in such nonlinear system; -- a chaos of both the solution and the structure!

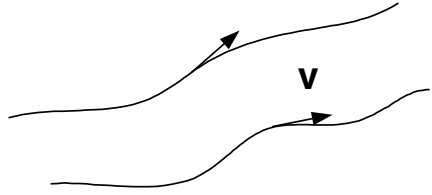

Fig. 7: A liquid flow. We see its "structure" directly in terms of the three components of **v** which are some of the state-variables. There is also the *pressure* that is not seen, and which, as the fourth variable requires the addition of one scalar equation (div **v** = 0) to the vectorial Navier-Stokes equation, allowing us to have 4 equations. However, it is sufficient to observe **v** ⊂ **x**, in order to speak about "**x**", and thus we pass on from (10) to (12).

Consider the Navier-Stokes equation ($\eta$ is viscosity, and $\rho$ density)

$$\frac{\partial \vec{v}}{\partial t} + (\vec{v}\nabla)\vec{v} = \frac{\eta}{\rho}\Delta\vec{v} + \frac{1}{\rho}\nabla p \qquad (10)$$

with the added condition for an incompressible liquid

$$\nabla\vec{v} = 0. \qquad (11)$$

Following [8] we omit in (10) the term containing the pressure (together with the now "redundant" (11)), obtaining

$$\frac{\partial \vec{v}}{\partial t} = -(\vec{v}\nabla)\vec{v} + \frac{\eta}{\rho}\Delta\vec{v} \qquad (12)$$

which can be rewritten (just introduce into the right-hand side the unit matrix [I], 3×3) in the spirit of (5a), as

$$\frac{\partial \vec{v}}{\partial t} = [A(\vec{v},\nabla)]\vec{v} \qquad (13)$$

with a nonlinear diagonal matrix-*operator*,

$$[A(\vec{v},\nabla)] = (-\vec{v}\nabla + \frac{\eta}{\rho}\Delta)[I]. \qquad (14)$$

See [4] for some more details, and especially the constructive results obtained in [8] using the system approach applied to the liquid flow.

It can be noted that though [8] in which (12) was studied is an absolutely independent work, one can see (12) as a 3D-generalization of the scalar *Burger's equation*

$$\frac{\partial u}{\partial t} + u \frac{\partial u}{\partial x} = k \frac{\partial^2 u}{\partial x^2} \qquad (15)$$

widely used (I did not know that when writing [8]) for demonstrating the specificity of some hydro- and aero-dynamic phenomena.

We conclude that some very basic physical situations are relevant to our outlook on the classification of system equations.

## 15. Some pedagogical philosophy

It is objectively not easy for one to combine riding the "highway" of a certain (e.g. chaos-theory) intensive modern studies with following the advice by R. P. Feynman [9] to make, from time to time a "step aside" from the established way of research and reconsider, or generalize, the basics of this way. However, the mentioned necessity to make system theory a part of one's general education requires equal rights to be given to the "bursting force" and "bases revisiting" approaches, and one has always remember that:

*Each time when a scientific discovery is done, it not only shows us a new perspective of scientific development, but also causes us to reconsider and better see what was done in the past* (F. Klein).

If the drivers on the fashion highways want to see how these highways are directed in the space of basic science, they should not forget about their responsibility of reconsidering the logical foundation of the theory/way each time when an advance in the field is achieved. In the present case, such an advance is the greatly increased role of switched (singular) systems in modern electronics, the consequent appearance of systems with numerous inputs, etc..

The classical pedagogical ways of introducing nonlinear systems should be changed, to include the concepts of singular "**t\***(**x**)-systems" and "**t\***(**u**)-systems", i.e. of "**t\***-nonlinearity". Many compliments must be paid to the position of [10,11] where nonlinear circuits are suggested for empirical study *before* linear circuits, and singular systems are the focus, i.e. the whole field of nonlinear systems is introduced, using the expression of Immanuel Kant, as "das Ding an sich" (a thing in/by itself), and *not* as something non-constructively defined as "*not a linear one*". (Indeed, *what is it then?*) The *constructive approach* to the very definition of nonlinearity is preferred for many reasons (work [4] discusses this in great detail), and, in particular, one sees that any use of the very term "system" already requires a *mathematical representation* of the system in focus to be given, from which it is already seen whether or the system is linear or nonlinear. One will not define (5) as "not (4)", as one will not define a curve as "not a straight line". *The concept of "given structure" has to become dominant in the definitions*.

The constructivism is thus required even by the axiomatic aspect. However this point is [4] somewhat philosophical and can lead us too far from the simple observations.

## 16. Uff! We learned a lot! Who invited this ugly duckling?!

1. Preferring, when possible, the "more structural" state-space [A(**x**)]-representation of some nonlinear systems, to the normal form representation d**x**/d$t$ = **F**($t$,**x**($t$),**u**($t$)), is a good line to thinking out some basic system points.

2.   The **t\***(.)-outlook is a very simple and heuristically very important outlook/principle: *watch which functions, -- known or unknown, -- define the switching instants* and thus define whether your switched system is NL or LTV.  It is really just the *understanding (observing)* that such an influence (a map) *exists*; no explicit analytical dependence involved in **t\***(**x**) is necessary for deriving conclusion re the very fact of nonlinearity.

3.   Switching nonlinearity appears "elegantly"; i.e. *not* via overstresses of some magnetic, ferroelectric, elastic, etc., materials, but *informatively*, through seeing that **x → t\***, and the *linear by themselves* switched elements need not be overstressed by either voltage or current for however strong nonlinearity to be obtained by means of the switchings.

4.  The fact, expressed by (3), that the definitions of LTV and NL (of both kinds, **x**- and **u**-) switched systems are close, is remarkable.  Where else are linear and nonlinear versions of a system defined in so analytically similarly terms?  However one should *not* (!) conclude from (3) that all switched systems are NL (or LTV).

5.  The role of the so-common-today switched systems in the logical foundation of system theory is noticeable.  In particular, the notations  **t\***($t$)-system, **t\***(**x**)-system, and **t\***(**u**)-system, or even simpler: "$t$-system", "**x**-system", and "**u**-system", are clear and suitable.

6.   There was no place here to consider, from the same positions, another very important singular operation, -- *sampling*.  Since such a consideration appears to be no less interesting and not less deep than that of switching, the reader is asked not to ignore the ArXiv references.

7.  Some non-trivial analogies where observed showing that the **t\***-nonlinearity, and the [A(**x**)]-system representation are relevant to some physical systems "as old as this world" which are around us.  The "system outlook" or "system thinking" has to become a part of one's perception of the world.  A more prosaic, though unexpected, example of such thinking can be the following biological hypothesis/question. *Can it be that at certain age, our body starts to (intentionally) develop processes destroying the teeth, if the amount of food that one usually eats is too large for the body's needs*?  That is, the destroying of teeth, which bothers eating, is a kind of *self regulation* (or a *feedback*) and it has to be understood by one as the requirement of the body to reduce his eating habits, and *not* as advice to immediately run to a dentist.

8.  The reader is asked to complete the argument of the second section re radiation of switched circuits by consideration of power losses, inevitable during quick charging of capacitors, or switchings between capacitors.  See also [4].

9.  The following homework is relevant to the very basics of our approach to switched circuits.  Consider that the principle of defining of $t_k(x_p)$ as some level-crossings of $x_p(t)$ contradicts the assumption of infinitely quick charging or discharging of capacitors, because an infinitely steep slope of a function involved makes $t_k$ non-sensitive to the crossing conditions.  Thus, some small resistances of the switches has to be introduced for the determination of $t_k$.  We think that the requirement of a realistic rate of a process is a good realistic/physical feature of the method of definition of $t_k(x_p)$, or of map (2), using level-crossings, and this (one more axiomatic) point can be demonstrated using the circuit of Fig 8.

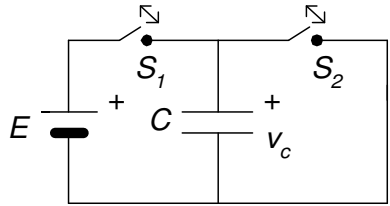

Fig. 8: The known circuit for modeling a linear resistor dependent on the operating frequency $f$ of the switches. By using some critical voltage level $v_{cr}$, at which $S_1$ is closed and $S_2$ open, we can make the circuit nonlinear just as the circuit in Fig. 5, but this requires the introduction of some small resistances of the switches, since processes described by δ-functions are obviously not appropriate for the definition (by means of comparators, as in Fig.1) of the instants of singularity as level-crossings; the $t_k$ will be non-sensitive to $v_{cr}$.

If the instants $t_k$ of closing $S_1$ (and opening $S_2$) are prescribed, then this circuit is LTV and it models, as is well known, the linear resistor $R = 1/fC$, defined in the sense of the average current $<i>$ taken from the source, $R \equiv E/<i>$. If $t_k$ is/are dependent on $v_c(t)$, then this circuit, and the modeled resistor, are nonlinear. However in order for the physical dependence of $t_k$ on $v_c(t)$ to be obtained/given by means of the using some $v_{cr}$, as for the circuit of Fig. 5, one has to introduce some small resistances of the switches. Then, this analysis is similar to that related to Fig. 5. Determination of the nonlinear resistive characteristic that the source "sees" when some $v_{cr}$ is thus employed, as well as analysis of any possible circuit variations, are left for the Reader.

## 17.  Open Problems/Questions

Our advice for a designer of switched systems is to define or watch the nonlinearity by defining (observing) the nature of the inputs of the comparators, this approach to become usual in the analysis of switched circuits. It would be useful to create a computer program which would *automatically check* whether or not a circuit is linear or nonlinear, meaning complicated structures, since for such simple structures, as, e.g., the known buck and boost power-electronics converters, the **t\*(x)**-nonlinearity can be seen immediately. Since our outlook is first of all logical, and only then analytical; i.e. we first have to just check *whether or not* some $x_p$ are the triggers; a *presentation of this outlook in terms of the circuits' graphs* should be useful. Thus, the first open problem is:

*Reformulate our point of view on the linearity and nonlinearity of switched systems using circuit-graph terms, and computerize the checking of the nature of* **t\***.

For the second point we refer to the example of [6] where the definition of [A(**x**)] is not done in terms of the zerocrossings of the $x_p(t)$ as we do. This different possibility suggests the general *classification* of switched systems in terms of different possible ways of obtaining [A(**x**)]:

*What are, qualitatively, the methods (possibilities) for obtaining* [A(**x**)] *and/or* [B(**x**)] *in switched systems?*

The third problem is:

*To find and classify different practical network schemes where the inputs are defined so that the whole system (or its subsystem) is **u**-nonlinear.*

The fourth problem (question) belongs to *sampling* systems that are considered in [3,4]. Here the situation regarding "linear or nonlinear" is quite similar to that of switched systems; however the points of singularity, **t***($t$) or **t***(**x**), are now the *sampling instants*. A mathematically basic point, raised in [4], is associated with the use of the level-crossings to create the approximating sums in the scheme of construction of the Lebesgue's integral. We compare the integral sums, those leading to Riemann's integral:

$$\sum_k f_k \, \Delta t_k \qquad (16)$$

where all $\Delta t_k$ are *independent* of $f(t)$, to those leading to Lebesgue's integral:

$$\sum_k f_k \cdot (\sum_m |\Delta t_{k,m}|) \qquad (17)$$

where the "measure"

$$\sum_m |\Delta t_{k,m}| = meas(f_k, \Delta f_k) \, ,$$

associated with $f_k$, is composed of $\Delta t_{k,m}$ defined by level-crossings of $f(t)$ by the two horizontal lines, $f = f_k - \Delta f_k /2$ and $f = f_k + \Delta f_k /2$. (Obviously, there can be several such $\Delta t_k$, for an $f(.)$ and an $f_k$, and we label these different '$\Delta t_k$' by $m$.)

That is, the time axis here is not independently "chopped" as in the Riemannian case, and the sampling instants, $t_k$, depend on $f(.)$. Thus, obviously, (17) is nonlinear by $f$, while (16) is linear, even though the very Lebesgue's *integral*, obtained from (17) in the limit of $\Delta f_k \to 0$, $\forall k$, is linear by '$f$' just as is Riemann's integral.

Thus, if the input function is unknown and we speak about signal analysis, i.e. $f(t)$ is a kind of $x(t)$, then the use of $t_k$ defined by level-crossings of $f(t)$ makes the system which realizes the finite Lebesgue's sums nonlinear in the **t***(**x**)-sense.

Sending the reader for a wider discussion to [4], let us formulate here our question as:

*Where can the nonlinearity of the finite Lebesgue's sums find engineering applications?*

The last question is the pedagogical one:

*What are the examples from physics, biology, sociology, etc., where the* A[**x**]- *(i.e. the structural) representation of nonlinearity is most natural, and what might be a good collection/set of such examples, taken from the different fields, which would be helpful in writing a popular book or textbook on system theory?*

This is an academic challenge, and such a pedagogical attempt would be in the spirit of the following, most correct, I think, definition of the goal of technical education:

*The goal of education is to turn some good science into a game for the pupil and thus help him make this science a part of his general education, i.e. a part of his systematic thinking and logical perception of this world.*